\documentclass[a4paper,twocolumn,english,prl,superscriptaddress,showpacs,longbibliography,fixfloat]{revtex4-1}
\usepackage[latin9]{inputenc}
\usepackage{amsmath}
\usepackage{amssymb}
\usepackage{graphicx}

\makeatletter

\pdfpageheight\paperheight
\pdfpagewidth\paperwidth

\usepackage{xcolor}
\definecolor{darkblue}{rgb}{0.1,0.2,0.6} 
\definecolor{lightblue}{rgb}{0.1,0.1,1.0}
\definecolor{darkred}{rgb}{0.8,0.1,0.2}
\usepackage[colorlinks,citecolor=lightblue,linkcolor=darkblue,urlcolor=lightblue] {hyperref}
\renewcommand{\BibitemShut}[1]{}

\makeatother

\usepackage{babel}
\begin{document}
\global\long\def\E{\mathrm{e}}
\global\long\def\D{\mathrm{d}}
\global\long\def\I{\mathrm{i}}
\global\long\def\mat#1{\mathsf{#1}}
\global\long\def\vec#1{\mathsf{#1}}
\global\long\def\cf{\textit{cf.}}
\global\long\def\ie{\textit{i.e.}}
\global\long\def\eg{\textit{e.g.}}
\global\long\def\vs{\textit{vs.}}
 \global\long\def\ket#1{\left|#1\right\rangle }

\global\long\def\etal{\textit{et al.}}
\global\long\def\tr{\text{Tr}\,}
 \global\long\def\im{\text{Im}\,}
 \global\long\def\re{\text{Re}\,}
 \global\long\def\bra#1{\left\langle #1\right|}
 \global\long\def\braket#1#2{\left.\left\langle #1\right|#2\right\rangle }
\global\long\def\obracket#1#2#3{\left\langle #1\right|#2\left|#3\right\rangle }
 \global\long\def\proj#1#2{\left.\left.\left|#1\right\rangle \right\langle #2\right|}

\title{Spin transport in long-range interacting one-dimensional chain}

\author{Benedikt Kloss }

\affiliation{Department of Chemistry, Columbia University, 3000 Broadway, New
York, New York 10027, USA}
\email{bk2576@columbia.edu}

\author{Yevgeny Bar Lev }

\affiliation{Department of Condensed Matter Physics, Weizmann Institute of Science,
Rehovot 76100, Israel}

\affiliation{Max-Planck-Institut für Physik komplexer Systeme, 01187 Dresden,
Germany}
\email{yevgeny.barlev@weizmann.ac.il}

\begin{abstract}
We numerically study spin transport and nonequilibrium spin-density
profiles in a clean one-dimensional spin-chain with long-range interactions,
decaying as a power-law, $r^{-\alpha}$ with distance. We find two
distinct regimes of transport: for $\alpha<1/2$, spin excitations
relax \emph{instantaneously} in the thermodynamic limit, and for $\alpha>1/2$,
spin transport combines both diffusive and superdiffusive features.
We show that while for $\alpha>3/2$ the spin diffusion coefficient
is finite, transport in the system is never strictly diffusive, contrary
to corresponding classical systems.
\end{abstract}
\maketitle
\emph{Introduction}.---Be it gravity, electromagnetic force or dipole-dipole
interactions, power-law interactions are ubiquitous. While sufficiently
dense mobile charges are able to screen the interaction and effectively
truncate its range, in many cases long-range interactions are important.
A few of the notable examples in conventional condensed matter systems
are nuclear spins \citep{Alvarez2015a}, dipole-dipole interactions
of vibrational modes \citep{Levitov1989,Levitov1990,Aleiner2011},
Frenkel excitons \citep{Agranovich2008}, nitrogen vacancy centers
in diamond \citep{Childress2006,Balasubramanian2009,Neumann2010,Weber2010,Dolde2013}
and polarons \citep{Alexandrov1996}. Long range interactions are
also common in atomic and molecular systems, where interaction can
be dipolar \citep{Saffman2010,Aikawa2012,Lu2012,Yan2013,Gunter2013,DePaz2013},
van der Waals like \citep{Saffman2010,Schauss2012}, or even of variable
range \citep{Britton2012,Islam2013,Richerme2014a,Jurcevic2014a}.

It was rigorously established by Lieb and Robinson that generic correlations
in quantum system with \emph{short-range} interactions propagate within
a linear ``light-cone'', $t/v=x$, with a finite velocity \citep{Lieb1972}.
Outside this ``light-cone'' correlations are exponentially suppressed
\citep{Lieb1972}. Specifically this implies that transport in local
quantum systems cannot be faster than ballistic. Lieb-Robinson bounds
were shown to be saturated for generic clean \citep{Bohrdt2016} and
weakly disordered systems \citep{Luitz2017}.

For systems with long-range interactions the result of Lieb and Robinson
doesn't hold, but was later generalized by Hastings and Koma, who
showed that for $\alpha>1$, the causal region in such systems becomes
at most logarithmic, $t\sim\log x$ \citep{Hastings2006}. This result
was subsequently improved to an algebraic ``light-cone'', $t\sim r^{\delta}$
for $\alpha>2$ and $0<\delta<1$ \citep{Foss-Feig2015}. A Hastings-Koma
type bound was also obtained for $\alpha<1$ after a proper rescaling
of time \citep{Storch2015}. While the spreading of generic correlations
was numerically studied in a number of studies \citep{Eisert2013,Hauke2013,Gong2014,Buyskikh2016,VanRegemortel2016,Luitz2018z},
much less is known about transport in long-range interacting systems.
Some information can be gained from quadratic fermionic models with
long-range hopping \citep{Maghrebi2016}, however these systems are
integrable and many times show nongeneric features. The results of
Ref.~\citep{Foss-Feig2015} suggest that transport in long-range
systems is \emph{at most} superdiffusive for $\alpha>2$, but leaves
a number of important questions open: (a) Is there an $\alpha$ above
which diffusion is recovered, similarly to the situation for classical
Lévy flights? \citep{Metzler2000} (b) Is there an $\alpha$, below
which mean-field like dynamical behavior takes place?

\begin{figure}
\includegraphics[width=86mm]{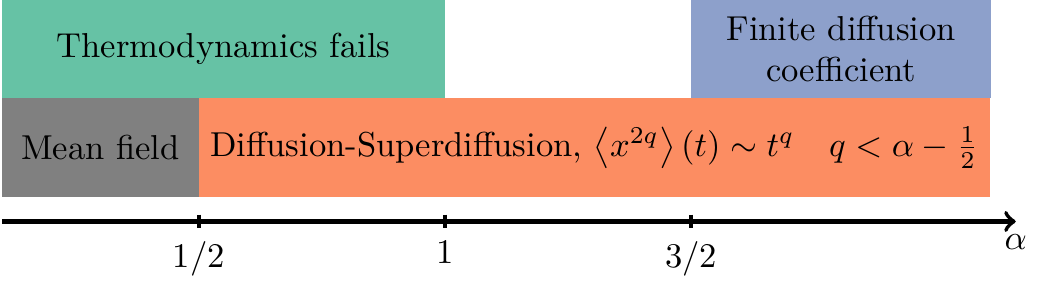}

\caption{\label{fig:cartoon}A cartoon describing the nature of transport in
one-dimensional interacting systems, with an interaction decreasing
as $r^{-\alpha}$ with the distance. For $0<\alpha<1$ the energy
of the system is superextensive, resulting in the failure of conventional
thermodynamics. For $0<\alpha<1/2$, dynamics corresponds to dynamics
of the infinite-range $\left(\alpha\to0\right)$ mean-field model
in the limit of $L\to\infty$. For $\alpha>1/2$ transport combines
diffusive and superdiffusive features.}
\end{figure}
In this work we address these questions using the time-dependent variational
principle in the manifold of matrix product states (TDVP-MPS) \citep{Verstraete2004,Haegeman2011,Haegeman2013,Haegeman2016}.
The main outcome of our study can be read from the cartoon in Fig.~\ref{fig:cartoon}.
TDVP-MPS belongs to the family of matrix product states (MPS) methods
\citep{Schollwock2011}, and thus allows to study long spin chains
(chains up to $L=1,601$ were considered here), way beyond what is
accessible using exact diagonalization. The main advantage of this
method over the conventional time-evolving block decimation (TEBD)
or time-dependent density matrix renormalization group (tDMRG) approaches
for time-evolution \citep{Vidal2004,White2004,Daley2004} is that
the evolution is unitary by construction, and the method explicitly
conserves a number of macroscopic quantities, such as the total energy,
total magnetization and total number of particles \citep{Verstraete2004,Haegeman2011,Haegeman2013,Haegeman2016}.
Moreover unlike TEBD and tDMRG the method can be directly applied
for long-range interacting systems. While the method is numerically
exact in the limit of large bond dimension (which sets the number
of variational parameters), it is limited by the growth of entanglement
entropy with time \citep{Schollwock2011}. For a fixed bond dimension,
the equations of motion of TDVP-MPS can be derived from a classical
nonquadratic Lagrangian in the space of variational parameters \citep{Haegeman2011,Leviatan2017}.
These equations are typically chaotic and yield diffusive transport.
Based on this observation as well as the conservation properties of
TDVP-MPS it was argued that the method could potentially recover correct
hydrodynamic behavior also for a relatively small bond dimension \citep{Leviatan2017},
a result which was challenged in Ref.~\citep{Kloss2017}. We note
in passing that this line of thought is not applicable for long-range
systems, where diffusive transport is not expected \emph{a-priori},
and the entire hydrodynamic approach is questionable. Therefore here
we strictly use TDVP-MPS as a numerically exact method.

\emph{Model}.---We study a one-dimensional spin-chain of length $L$,
given by the Hamiltonian $\hat{H}=\hat{H}_{\text{loc}}+\hat{H}_{\text{lr}}$
where
\begin{equation}
\hat{H}_{\text{loc}}=\sum_{i=1}^{L-1}\left(\hat{S}_{i}^{x}\hat{S}_{i+1}^{x}+\hat{S}_{j}^{y}\hat{S}_{i+1}^{y}\right)+\sum_{i=1}^{L-2}\left(\hat{S}_{i}^{x}\hat{S}_{i+2}^{x}+\hat{S}_{j}^{y}\hat{S}_{i+2}^{y}\right),
\end{equation}
is the local part and,
\begin{multline}
\hat{H}_{\text{lr}}=\sum_{i=1}^{L-1}\sum_{j>i+1}^{L}\frac{1}{\left(j-i-1\right)^{\alpha}}\left(\hat{S}_{i}^{x}\hat{S}_{j}^{x}+\hat{S}_{i}^{y}\hat{S}_{j}^{y}\right),\label{eq:LR_XX_H}
\end{multline}
includes power-law decaying long-range interactions and $\hat{S}_{i}^{x}$
and $\hat{S}_{i}^{y}$ are spin-1/2 operators. The Hamiltonian conserves
the total magnetization, and thus supports energy and spin transport.
In the limit of $\alpha\to\infty$, $\hat{H}_{\text{lr}}$ vanishes,
and the resulting Hamiltonian corresponds to the XX ladder, which
is nonintegrable and has diffusive spin transport \citep{Steinigeweg2014,Kloss2017}.

\begin{figure}
\includegraphics{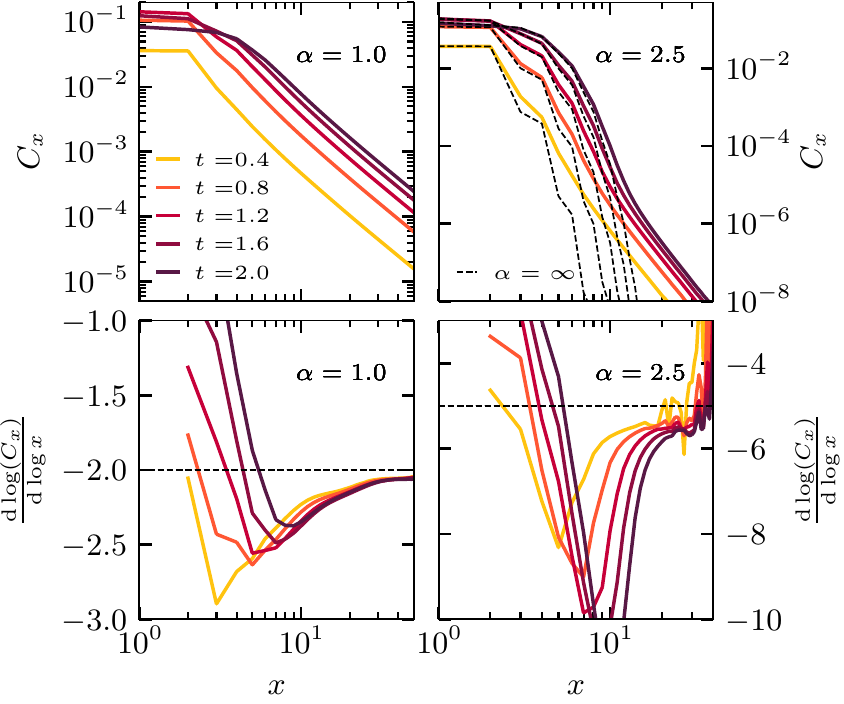}

\caption{\label{fig:profile-t}\emph{Upper panels}. Spin excitation profiles
as a function of time for two representative $\alpha$. The dashed
black lines correspond to results obtained in the $\alpha\to\infty$.
Darker tones represent longer times. \emph{Lower panels}. Logarithmic
derivative of the spin excitation profiles. The dashed black lines
are guides to the eye for $2\alpha$. $L=201$, $\chi=256$ .}
\end{figure}
\emph{Method}.---To assess spin transport in the system we numerically
compute the two-point spin-spin correlation function at infinite temperature,
\begin{equation}
C_{x}\left(t\right)=\frac{4}{2^{L}}\tr\hat{S}_{L/2+x}^{z}\left(t\right)\hat{S}_{L/2}^{z}\left(0\right),
\end{equation}
which corresponds to the time-dependent profile of a local excitation
at the center of the chain performed at $t=0$. The excitation profile
is obtained by propagating the operators, $\hat{S}_{i}^{z}\left(t\right),$under
the Heisenberg evolution. Accessible timescales are limited by the
growth of entanglement entropy during the time-evolution. Using the
cyclic property of the trace $C_{x}\left(t\right)$ can be written
as,
\begin{equation}
C_{x}\left(t\right)=\frac{4}{2^{L}}\tr\hat{S}_{L/2+x}^{z}\left(-\frac{t}{2}\right)\hat{S}_{L/2}^{z}\left(\frac{t}{2}\right),
\end{equation}
which allows us to reach twice as large times \citep{Kennes2014}.
Since we work with an approximately translationally invariant system
(we use open boundary conditions), in practice, we propagate only
\emph{one} operator at the center of the lattice, since operators
which are far enough from the boundaries of the chain can be obtained
approximately by a simple translation \citep{SuppMat2018}. To mitigate
the boundary effects introduced by this approximation we show $C_{x}\left(t\right)$
only for the central $L/2$ sites of the chain. If not stated otherwise,
we use spin-chains of length $L=201$, which is sufficient to have
finite size effects under control for most ranges of the interaction. 

To propagate the operators we use the time-dependent variational principle
(TDVP), which yields a locally optimal (in time) evolution of the
wavefunction on some variational manifold. It amounts to solving a
tangent-space projected Schrödinger equation \citep{Haegeman2016},
\begin{equation}
\frac{d}{dt}\ket{\hat{O}\left(t\right)}=-\I P_{\mathcal{M}}\hat{H}\ket{\hat{O}\left(t\right)},\label{eq:TDVP}
\end{equation}
where $P_{\mathcal{M}}$ is the tangent space projector to the variational
manifold $\mathcal{M}$ and $\ket{\hat{O}\left(t\right)}$ is a vectorization
of a general operator $\hat{O}\left(t\right)$. We use the matrix
product operator (MPO) representation of the operator,
\begin{multline}
\hat{O}\left(\boldsymbol{A}\right)=\sum_{\left\{ \sigma_{i}\right\} ,\left\{ \sigma_{i}'\right\} }A_{1}^{\sigma_{1}\sigma_{1}'}\dots A_{N}^{\sigma_{N}\sigma_{N}'}\ket{\sigma_{1}\dots\sigma_{n}}\bra{\sigma_{1}'\dots\sigma_{n}'},\label{eq:MPO_Gen}
\end{multline}
where $\sigma_{i}=\pm1/2$ correspond to the states of a spin at site
$i$ and $A_{i}^{\sigma_{i}\sigma_{i}'}\in\mathbb{C}^{\chi_{i-1}\times\chi_{i}}$
are complex matrices where $\chi_{i}$ is the bond-dimension of the
matrix $\left(\chi_{0}=\chi_{N}=1\right)$ \citep{Schollwock2011}.
An exact representation of a general operator requires the bond dimension
to grow exponentially with system size $L$. Therefore truncating
the maximal bond-dimension to a fixed value introduces an approximation
but allows to keep the MPO representation tractable. We use the family
of fixed finite bond-dimension MPOs to parameterize the variational
manifold, $\mathcal{M}$. Numerically exact results are achieved by
convergence with respect to the bond-dimension (in this work we used
bond-dimension of up to 256) \citep{SuppMat2018}. The evolution of
(\ref{eq:TDVP}) is performed using a second-order Trotter decomposition,
with time-steps from 0.01 to 0.05. The Hamiltonian is approximated
as a sum of exponentials and a short-ranged correction, which can
be efficiently represented as an MPO. The number of exponentials is
chosen such that the resulting couplings do not differ more than 2\%
from the exact couplings for any pair of sites \citep{Crosswhite2008a}.
We note in passing that since the evolution is unitary in the \emph{enlarged}
vector space of the vectorized operators and the method explicitly
conserves the norm of the operator, $\braket{\hat{O}\left(t\right)}{\hat{O}\left(t\right)}\equiv\tr\hat{O}^{\dagger}\left(t\right)\hat{O}\left(t\right)=\tr\hat{O}^{\dagger}\hat{O}$,
but \emph{not }its trace, $\tr\hat{O}\left(t\right)$.

\begin{figure}
\includegraphics{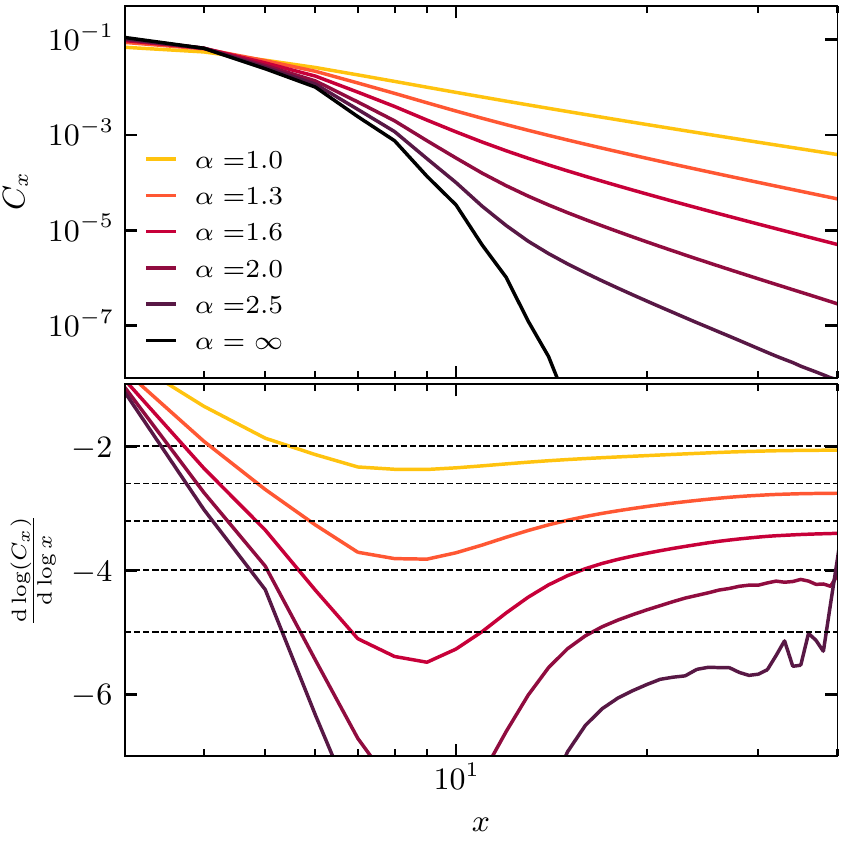}

\caption{\label{fig:profile}\emph{Upper panel}. Spin excitation profiles at
$t=2.0$ and various $\alpha$. Darker tones represent larger $\alpha$-s.
\emph{Bottom panel}. Logarithmic derivative of the spin excitation
profiles. The dashed black lines are guides to the eye for $2\alpha$.
$L=101$, $\chi=256$.}
\end{figure}
\emph{Results}.---Figure~\ref{fig:profile-t} shows the spin excitation
profile, $C_{x}\left(t\right)$, for short times and two values of
$\alpha=1$ and $2.5.$ Since the excitation profile is symmetric
with respect to the center of the lattice in the following figures
we only show its right side $\left(x>0\right)$. For $\alpha=2.5$,
and small distances from the initial excitation, the profile resembles
a Gaussian and superimposes well with the $\alpha\to\infty$ profile
calculated at same time points. For larger distances there is a crossover
from a Gaussian form to a power-law form, $x^{-\gamma}$, which becomes
increasingly pronounced as the time progresses. For smaller $\alpha$,
the crossover is less pronounced and there is no apparent region of
Gaussian behavior (although it might develop at later times). Since
the accessible times in this work are short $\left(t\leq4\right),$
due to fast growth of entanglement entropy, it is pertinent to question
what our results imply on bulk transport? From Fig.~\ref{fig:profile-t}
it is apparent that the power-law tail appears already at very short
times, and its exponent $\gamma$ seems to be independent of time,
as can be judged from convergence of the logarithmic derivative, $\mathrm{d}\log C_{x}\left(t\right)/\mathrm{d\log x}$,
to the same value of $\gamma$ (see bottom panels of Fig.~\ref{fig:profile-t}).
This leads us to argue that the long-range nature of the interactions
speeds up the approach to asymptotic transport and allows us to observe
at least some of its features.

In Fig.~\ref{fig:profile} we show the spin excitation profile at
$t=2$ for all analyzed $\alpha$. The power-law regime, $x^{-\gamma}$,
is visible for all $\alpha$ and the exponent $\gamma\left(\alpha\right)$
is $\alpha$ dependent. To assess this dependence we calculate the
corresponding logarithmic derivative (see bottom panel of Fig.~\ref{fig:profile}),
which converges to its asymptotic value, $\gamma$, at large distances.
The logarithmic-derivative becomes increasingly noisy at large distances,
$x$, (where $C_{x}\left(t\right)<10^{-8}$ ), due to decreasing signal-to-noise
ratio, which prohibits us to obtain an even better convergence.

\begin{figure}
\includegraphics{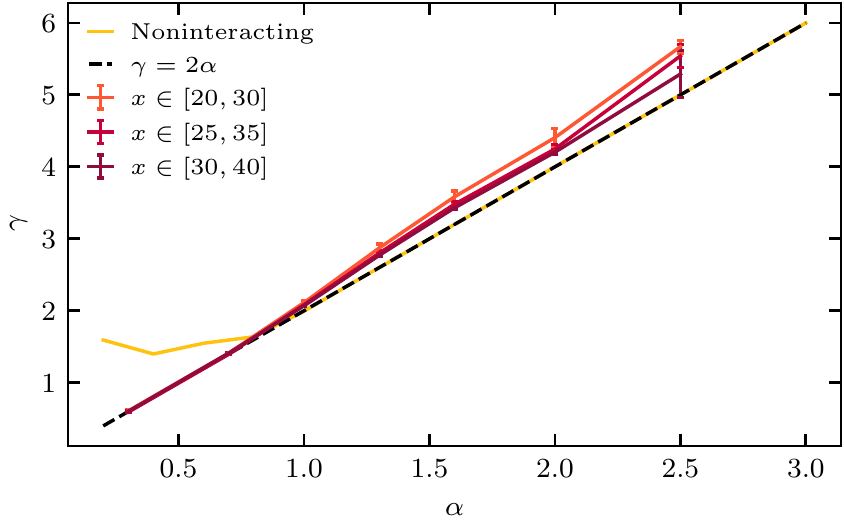}

\caption{\label{fig:exp}The power-law exponent, $\gamma$, of the power-law
tail in the spin excitation profiles obtained by averaging over the
logarithmic derivative in Fig.~\ref{fig:profile} in different spatial
regions. The yellow (light) line is the exponent $\gamma$ computed
for the noninteracting model in Eq.~(\ref{eq:nonint}). The dashed
black line corresponds to $\gamma=2\alpha$.}
\end{figure}
The Hastings-Koma bound (as also the tightened algebraic bounds) states
that for $\alpha>1$, $\left\Vert \left[\hat{O}_{y}\left(t\right),\hat{O}_{x+y}\right]\right\Vert \leq c_{\alpha}\left(t\right)x^{-\alpha}$,
where $\hat{O}_{i}$ are generic local operators, and $c_{\alpha}\left(t\right)$
is a constant, which depends on $t$ and $\alpha$ \citep{Hastings2006}.
Since $C_{x}\left(t\right)$ is a correlation function, one could
expect the exponent of the power-law decay of $C_{x}\left(t\right)$
to simply be $\alpha,$ namely $\gamma=\alpha$. This is however not
the case as can be inferred from Fig.~\ref{fig:exp}. To assess the
convergence of the results we have extracted $\gamma$ by averaging
the logarithmic derivative on various spatial intervals, and we note
that $\gamma$ converges to the $\gamma=2\alpha$ line. We compare
our results to a noninteracting long-range hopping model,
\begin{equation}
\hat{H}_{\text{nonint}}=\sum_{i=1}^{L}\sum_{x=1}^{L-1}\frac{1}{x^{\alpha}}\hat{c}_{i}^{\dagger}\hat{c}_{i+x},\label{eq:nonint}
\end{equation}
where $\hat{c}_{i}^{\dagger}$ creates a spinless fermion at site
$i$ (for analytical results at the groundstate, see Refs.~\citep{Vodola2014,Vodola2015,Maghrebi2016}).
Interestingly, while both models yield similar results for $\alpha>1$,
they differ for $\alpha<1$, where the interacting model continues
to follow the $\gamma=2\alpha$ line.

\begin{figure}
\includegraphics{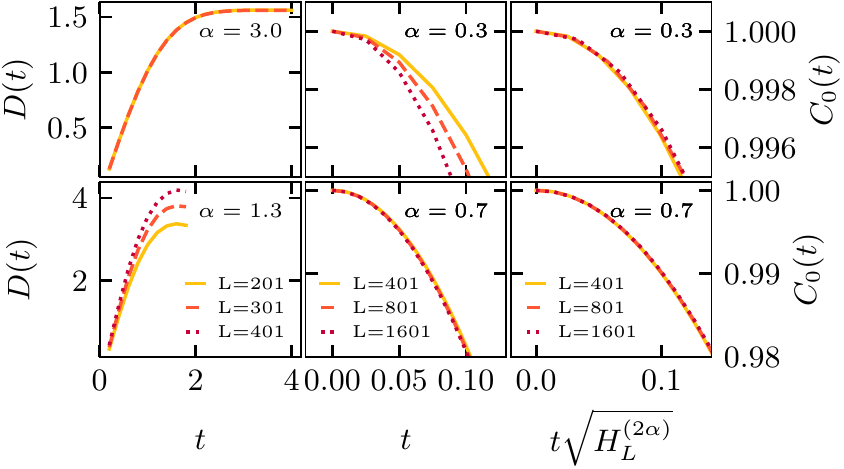}

\caption{\label{fig:MSD_rescale}\emph{Left panels}. Time-dependent diffusion
constant $D\left(t\right)$ for $\alpha=1.3$ and $3$ and three different
system sizes, $L=201,301$ and $401$, $\chi=256$. \emph{Middle and
right panels}. Short-time relaxation of the central spin, $C_{0}\left(t\right)$
for $\alpha=0.3$ ($C_{0}\left(5t\right)$ is plotted) and $0.7$
versus time (middle) and rescaled time (right) using the square root
of the generalized harmonic numbers, $\sqrt{H_{L}^{\left(2\alpha\right)}}$
(see text) for $L=401,801$ and $1,601$ and $\chi=64$.}
\end{figure}

Many-times transport is characterized by considering the time-dependence
of the moments of of the spin excitation profile,
\begin{equation}
\left\langle x^{2q}\right\rangle \left(t\right)=\sum_{x=0}^{L}x^{2q}C_{x}\left(t\right).\label{eq:MSD}
\end{equation}
Specifically, the second moment $(q=1),$ also known as the mean-square
displacement (MSD), is directly related to the the time-dependent
diffusion coefficient , $D\left(t\right)=\mathrm{d}\left\langle x^{2}\right\rangle /\mathrm{d}t$,
which converges to the linear response diffusion coefficient for $t\to\infty$
(see Appendix of Ref.~\citep{Luitz2016c}). Since we obtain that
asymptotically $C_{x}\left(t\right)\sim x^{-2\alpha}$, all moments
with $q>\alpha-1/2$ diverge in the limit $L\to\infty$. In the left
panels of Fig.~\ref{fig:MSD_rescale} we demonstrate this behavior
for $q=1$. . While $\alpha=1.3$ shows a divergence of $D\left(t\right)$
with system size, for $\alpha=3$ the time-dependent diffusion coefficient
does \emph{not} depend on the system size, and approaches a plateau
as a function of time, indicative of diffusive transport, $\left\langle x^{2}\right\rangle \sim Dt$.
This is consistent with our observation that the central part of the
excitation profile is well described by the dynamics of a local system
$\left(\alpha\to\infty\right)$, which is diffusive \citep{Steinigeweg2014,Kloss2017}.

We note that $\alpha=1/2$ plays a special role, since for $\alpha<1/2$,
$C_{x}\left(t\right)\sim x^{-2\alpha}$ becomes nonintegrable. This
is in a contradiction to the fact that $\sum_{x}C_{x}\left(t\right)=\sum_{x}C_{x}\left(0\right)=1$,
which follows from the conservation of total magnetization. The resolution
of this apparent paradox follows from the dependence of $C_{x}\left(t\right)$
on the system size for $\alpha<1/2$, which makes the entire excitation
profile (for any finite time) vanish in the limit $L\to\infty$ \citep{Bachelard2013,Kastner2011,VanDenWorm2013,Kastner2017}.
The dependence of the excitation profile on the system size for $\alpha<1/2$
can be eliminated by a proper rescaling of time, $tf\left(L\right),$
where $f\left(L\right)$ is some increasing function of $L$. We have
empirically found that taking $f\left(L\right)=\sqrt{H_{L}^{\left(2\alpha\right)}}\equiv\left(\sum_{x=1}^{L}x^{-2\alpha}\right)^{1/2}$
(namely the $\ell^{2}$-norm of the long-range part) gives a perfect
scaling collapse (see right panels of Fig.~\ref{fig:MSD_rescale})
for $\alpha<1$\footnote{For $1/2<\alpha$ such a rescaling is not required for sufficiently
large system sizes. For small system sizes a notable residual dependence
on the system size might exist, especially for $\alpha$ close to
$1/2$, due to slow convergence of $H_{L}^{\left(2\alpha\right)}$.
The same rescaling of time as we use for $\alpha<1/2$ eliminates
this residual dependence.}. In the limit of large system sizes and for $\alpha<1/2$, this rescaling
corresponds to $\tau\sim tL^{1/2-\alpha}$ and is consistent with
the analytically obtained rescaling for a classical model \citep{Bachelard2013,Kastner2017}.

\emph{Summary}.---Using a numerically exact method (TDVP) we study
spin transport in a nonintegrable one-dimensional spin chain, with
interactions which decay as $x^{-\alpha}$ with the distance. While
the method allows us to address chains far beyond what is accessible
using exact diagonalization, it is inherently limited to short times
due to the fast growth of entanglement entropy. Nevertheless, we show,
that due to the long-range of the interactions, approach to some of
the asymptotic features of transport is fast enough to be observed
in our simulations.

We find two pronounced regimes in the dynamics of a spin excitation.
For $\alpha<1/2$, we find that the decay of the excitation depends
on the system size, such that the relaxation time $t_{0}\propto\left(\sum_{k}J_{0k}^{2}\right)^{-1/2}\sim L^{\alpha-1/2}$
(where $J_{ij}\sim\left|i-j\right|^{-\alpha}$ is the long-range part
of the Hamiltonian), and goes to zero in the limit of $L\to\infty$.
For \emph{finite} system sizes the spatial decay of the excitation
profile is $C_{x}\left(t\right)\sim x^{-2\alpha}$.

For $\alpha>1/2$, there is a residual dependence of the excitation
profiles on the system size, which vanishes in the $L\to\infty$ limit.
For short distances the spatial excitation profiles are well described
by the corresponding profiles of a local system , which for generic
systems are Gaussian, corresponding to a diffusive transport. For
longer distances the Gaussian form crosses-over to a power-law behavior
with an exponent, which approaches, $C_{x}\left(t\right)\sim x^{-2\alpha}$.
The crossover is much more apparent for larger $\alpha$, and is barely
visible for the smaller $\alpha$. Our data is inconclusive with respect
to the existence of a critical $\alpha_{c}>1/2$ below which the crossover
vanishes, since it is possible that longer times are needed to observe
the crossover for the smaller $\alpha$. The crossover point drifts
to longer distances for larger $\alpha$, but we were not able to
determine its precise functional dependence.

Due to the asymptotic power-law dependence of the excitation profile,
only moments $\left\langle x^{2q}\right\rangle \left(t\right)$ with
$q<\alpha-1/2$ exist (see Eq.~\ref{eq:MSD}). We find that for $\alpha>3/2$
the MSD, which corresponds to $q=1$, exists and is \emph{not} system-size
dependent. Moreover it appears to increase linearly with time, which
we demonstrated by calculating its derivative. While this behavior
corresponds to diffusion, the dynamics is \emph{not} truly diffusive
for any $\alpha$, due to the divergence of higher moments. This is
in stark contrast to \emph{classical} superdiffusive systems, such
as Lévy flights, where a critical $\alpha$ exists, above which diffusion
is restored. The nice agreement of the ``core'' of the excitation
profile with a Gaussian form, corresponding to diffusion, leads us
to speculate that \emph{all} the existing moments have a diffusive
time-dependence, namely, $\left\langle x^{2q}\right\rangle \left(t\right)\sim t^{q}$,
for $q<\alpha-1/2$.

In this work we consider only one model, but due to its nonintegrability
for all $\alpha$, we expect our results to hold for a broad family
of nonintegrable long-range models. In particular, it would be interesting
to extend our results to higher dimensions.
\begin{acknowledgments}
BK acknowledges funding through the Edith \& Eugene Blout Fellowship.
This work used the Extreme Science and Engineering Discovery Environment
(XSEDE), which is supported by National Science Foundation Grant No.
OCI-1053575. 
\end{acknowledgments}

\bibliographystyle{apsrev4-1}
\bibliography{lib_yevgeny,local}

\cleardoublepage{}

\section*{Supplementary material}

\begin{figure}
\includegraphics[width=1\columnwidth]{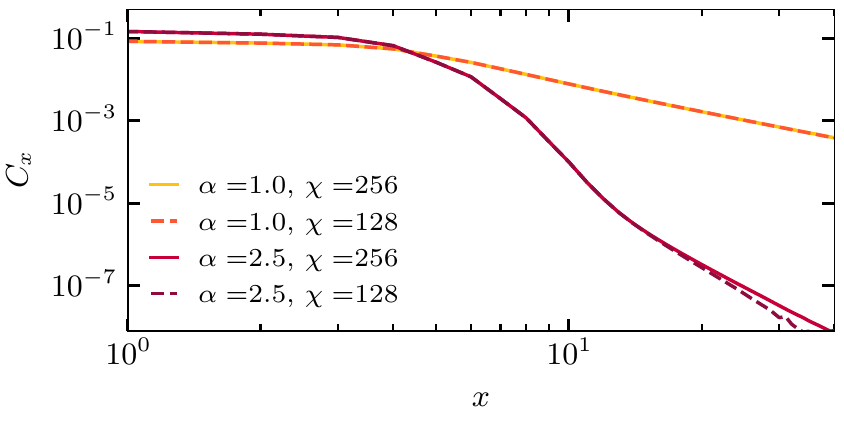}

\caption{\label{fig:conv-profile}Convergence of the spin excitation profile
with respect to bond-dimension, $\chi$, at $t=2.0$ and $L=201$.}
\end{figure}

\begin{figure}
\includegraphics[width=1\columnwidth]{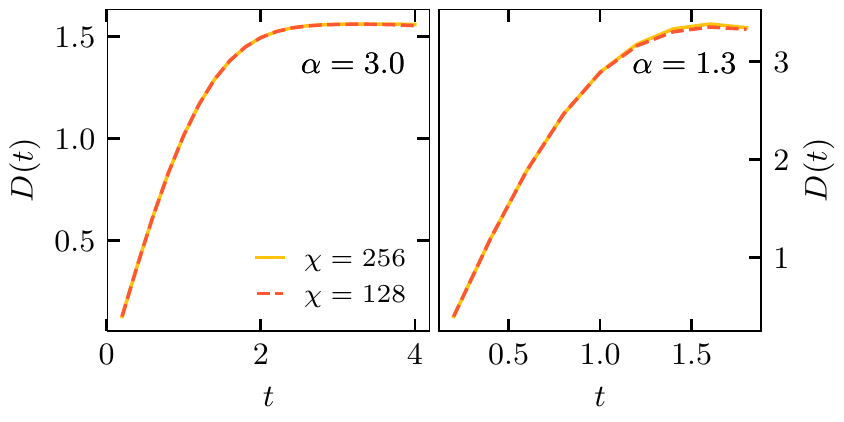}

\caption{\label{fig:conv-Dt}Convergence of the time-dependent diffusion constant
$D(t)$ with respect to bond-dimension for $\alpha=1.3$ $\left(L=201\right)$
and $\alpha=3$ $\left(L=301\right)$.}
\end{figure}

\begin{figure}
\includegraphics[width=1\columnwidth]{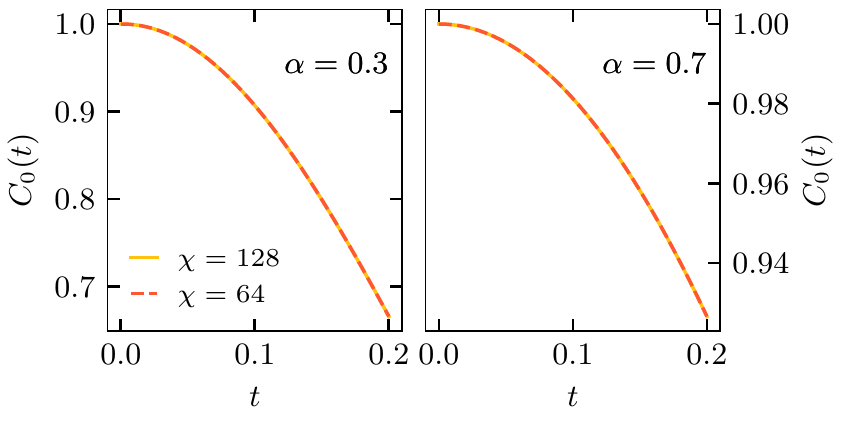}

\caption{\label{fig:conv-C0}Convergence of the relaxation of the central spin
$C_{0}(t)$ with respect to bond-dimension for $\alpha=0.3$ and $\alpha=0.7$
$\left(L=401\right)$.}
\end{figure}

\emph{Convergence tests}.---Numerical exactness of the dynamics generated
by TDVP-MPS is obtained by converging with respect to the bond-dimension,
$\chi$. In Figures \ref{fig:conv-profile} and \ref{fig:conv-Dt},
we provide comparisons of calculations with bond-dimensions $\chi=256$
and $\chi=128$ for quantities of interest in this study. Evaluating
the spatial spin excitation profile in the tails becomes sensitive
to numerical noise for small values of $C_{x}$ (smaller than $10^{-8}$)
and is limited by a complex interplay of time-step errors and accumulation
of numerical round-off errors. Therefore, obtaining accurate tails
of $C_{x}$ is harder for the large $\alpha$, where $C_{x}$ decreases
faster with the distance. $\alpha=2.5$ is the shortest-ranged system
for which it is possible to calculate a meaningful tail of $C_{x}$.
In contrast, the mean square displacement is robust to the numerical
noise in the far tails for the system sizes and times considered here,
and longer times are accessible for larger $\alpha$. The relaxation
of the central spin, $C_{0}(t)$, at short times is converged with
a moderate bond-dimension $\chi=64$, see Fig.~\ref{fig:conv-C0}.

\begin{figure}
\includegraphics[width=1\columnwidth]{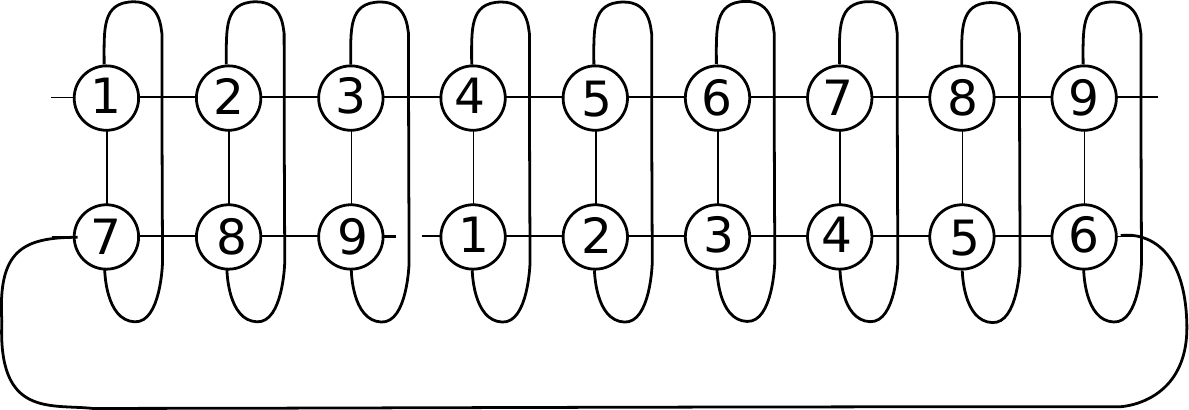}

\caption{\label{fig:MPO-cartoon}Tensor network diagram for Eq.~(\ref{eq:Profile_appr}).
Each tensor in the network is labeled with the physical site it represents.
The upper MPO corresponds to the untranslated operator $\hat{S}_{L/2}^{z}\left(-\frac{t}{2}\right)$
while the lower MPO is its translated version $\boldsymbol{T}_{3}\hat{S}_{L/2}^{z}\left(\frac{t}{2}\right)$.}
\end{figure}

\emph{Approximate evaluation of} $C_{x}\left(t\right)$.---Obtaining
the correlation function, 
\begin{equation}
C_{x}(t)=\frac{1}{2^{L}}\tr\hat{S}_{L/2}^{z}\left(-\frac{t}{2}\right)\hat{S}_{L/2+x}^{z}\left(\frac{t}{2}\right),
\end{equation}
of a spin-chain of length $L$ scales as $\mathcal{O}(L^{2})$, since
for each operator, a separate calculation has to be performed. However,
the scaling can be reduced to $\mathcal{O}(N)$ by making use of the
approximate translational invariance of the $\hat{S}_{i}^{z}(t)$.
In the limit of large system and for sites $i$ close to the center,
the correlation function can be evaluated approximately using only
$\hat{S}_{L/2}^{z}(t)$,
\begin{equation}
C_{x}\left(t\right)\approx\frac{1}{2^{L}}\tr\hat{S}_{L/2}^{z}\left(-\frac{t}{2}\right)\boldsymbol{T}_{x}\hat{S}_{L/2}^{z}\left(\frac{t}{2}\right),\label{eq:Profile_appr}
\end{equation}
where the action of the translation operator $\boldsymbol{T}_{x}$
is illustrated in Fig.~\ref{fig:MPO-cartoon}. It can be understood
as relabeling of the lattice sites $i$ in a cyclically translated
manner: $\forall i\in[1,L]:i\to(i+x)\mod L$. The trace in Eq.~(\ref{eq:Profile_appr})
can be performed if the matrix product operator (MPO) is expanded
at both ends with virtual sites connected containing identity operators
and connected with a bond-dimension of 1. One may also include these
sites as physical sites in the propagation, which corresponds to a
mean-field description of the auxiliary sites. In this study we chose
the latter and included $\frac{L}{2}$ auxiliary sites to the left
and right of the chain. The system sizes $L$ reported in the main
text refer to the lattice without the auxiliary sites. There is no
need to evaluate $\hat{S}_{L/2}^{z}\left(-\frac{t}{2}\right)$, since
it is just the complex conjugate of $\hat{S}_{L/2}^{z}\left(\frac{t}{2}\right)$.
In a vectorized notation the calculation of $C_{x}\left(t\right)$
therefore amounts to the calculation of $\left\langle \hat{S}_{L/2}^{z}\left(\frac{t}{2}\right)|\boldsymbol{T}_{x}|\hat{S}_{L/2}^{z}\left(\frac{t}{2}\right)\right\rangle .$

\begin{figure}
\includegraphics[width=1\columnwidth]{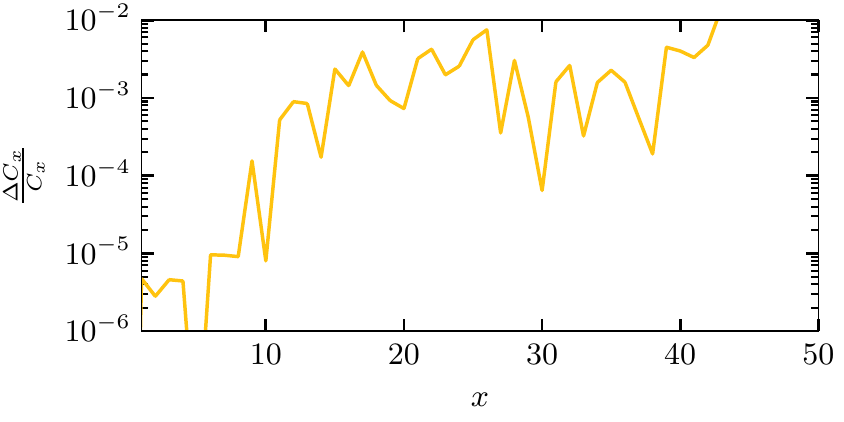}

\caption{\label{fig:explicit-vs-implicit}Relative deviation between the spin
excitation profiles obtained with and without the approximation described
in the text. Data shown is for $\mathrm{t=2.0,\thinspace\alpha=2.0,\thinspace dt=0.1,\thinspace\chi=128,\thinspace L=201.}$}
\end{figure}

The deviation between $C_{x}(t)$ obtained from the explicit propagation
of all $\hat{S}_{x}^{z}$ and $C_{x}(t)$ calculated within this approximation
is negligible for the chain lengths we use in this study, see Fig.~\ref{fig:explicit-vs-implicit}.
We have verified that the large errors after site 40 are not related
to a breakdown of the approximate scheme, but occur due to the small
signal-to-noise ratio for very small $C_{x}(t)$. For lattice sites
close to the end of the chain, the approximation is expected to cause
significant errors.

\end{document}